\documentstyle[revtex]{aps}

\begin{document}

\begin{title}
{Proof of the Thin Sandwich Conjecture}
\end{title}
\author{Robert Bartnik}
\begin{instit}
Department of Mathematics, Statistics and Computing Science\\
University of New England\\
Armidale NSW 2351, Australia
\end{instit}
\author{Gyula Fodor\cite{GF-Budapest}}
\begin{instit}
Centre for Mathematics and its Applications \\
Australian National University\\
Canberra ACT 2601, Australia
\end{instit}
\receipt{2nd April 1993}

\newtheorem{theo}{Theorem}
\newtheorem{lemma}{Lemma}
\newtheorem{assu}{Assumption}
\newtheorem{coro}{Corollary}
\newtheorem{propo}{Proposition}

\begin{abstract}
We prove that
the Thin Sandwich Conjecture in general
relativity is valid, provided that the data
$(g_{ab},\dot g_{ab})$ satisfy certain geometric conditions.
These conditions define an open set in the class of possible data,
but are not generically satisfied.
The implications for the ``superspace'' picture of the Einstein
evolution equations are discussed.
\end{abstract}
  \draft
\narrowtext

The ``superspace'' picture of the Einstein equations espoused by J.~A.~Wheeler
and others \cite{MTW,ChrFra}, envisages a spacetime as a curve in superspace
${\cal S}$, the space of Riemannian metrics on a fixed 3-dimensional manifold
${\cal M}$ \cite{comment}.
It was hoped that a useful analogy might be drawn with the classical dynamics
of a point particle and its quantum mechanical generalisation using the
Feynman path integral method, and with this objective Wheeler advanced the
(Thick) Sandwich and Thin Sandwich conjectures. The full Sandwich
Conjecture proposes that, given two 3-metrics $g_1, g_2\in{\cal S}$, there
exists a unique spacetime ${\cal V}$ in which  $g_1$, $g_2$ arise as the
induced
metrics on (disjoint) Cauchy hypersurfaces in ${\cal V}$ (``${\cal V}$ is a
spacetime
connecting $g_1$, $g_2$''). Foliating ${\cal V}$ between these Cauchy surfaces
determines a curve in ${\cal S}$ connecting $g_1$, $g_2$, but perturbing the
foliation changes the curve, so the local uniqueness of the classical
connecting path that holds in particle dynamics is not valid here. The
uniqueness assertion of the Sandwich Conjecture implies that this
diffeomorphism degeneracy is the only cause of non-uniqueness of the Einstein
flow in ${\cal S}$.

However, the electromagnetic analogue of the Sandwich Conjecture has an
ultraviolet degeneracy, with non-trivial non-uniqueness arising from
high-frequency components
--- there are (spatially periodic) non-trivial solutions of Maxwell's
equations having vector potential $A_\mu$ vanishing at two times $t=0,1$
\cite{Fodor}.
This implies that uniqueness must fail for solutions of the
Sandwich Conjecture for Maxwell's equations,
 so it is possible that the full conjecture for the
Einstein evolution may also be false. Such problems do not arise with the
Thin Sandwich Conjecture, which is analogous to the initial value problem in
particle dynamics and asserts that given a point $g\in{\cal S}$ and a tangent
vector $\dot g\in T_g{\cal S}$, there is a unique spacetime ${\cal V}$
realising the
initial condition $(g,\dot g)$. This means that ${\cal V}$ has a time function
$t$
and a time flow vector $T$ such that the metric in adapted coordinates $(t,x)$
has the form
\begin{equation}
ds^2=-N^2dt^2+g_{ij}(dx^i+X^idt)(dx^j+X^jdt), \label{metric}
\end{equation}
where $g_{ij}(x,t)$ is the induced metric on the level sets of $t$, $N$ is
the lapse and $X^i$ is the shift,
$T=\partial/\partial t=\partial_t$, and
\begin{equation}
g_{ij}=g_{ij}(0),\qquad
\dot g_{ij}=\frac\partial{\partial t}g_{ij}(0).
\end{equation}
The local existence and uniqueness theorems for the Einstein equations
\cite{ChoBru,ChBrGe} reduce this problem to one of constructing geometric
initial data $(g_{ij},K_{ij})$ satisfying the constraint equations
\begin{eqnarray}  \label{enco}
16\pi T_{00}\equiv 2\epsilon =R_g-K_{ij}K^{ij}+(K^k_k)^2,\\
8\pi T_{0i}\equiv S^i        =\nabla^j K_{ij}-\nabla_i K^j_j,\label{imco}
\end{eqnarray}
and such that the extrinsic curvature $K_{ij}$ is compatible with
$(g_{ij},\dot g_{ij})$ through the metric (\ref{metric}), ie.~there is
$(N,X^i)$ such that
\begin{equation}
K_{ij}=N^{-1}\left(\case1/2\dot g_{ij}-X_{(i|j)}\right).\label{excur}
\end{equation}
Substituting (\ref{excur}) into (\ref{enco},\ref{imco}) gives the Thin
Sandwich Equations (TSE), a system of differential equations for $(N,X^i)$,
where $(g_{ij},\dot g_{ij})$ and the (normalised) local energy and momentum
densities $(\epsilon,S^i)$ are given fields. Formally solving (\ref{enco})
for $N$
\begin{equation}
N=\left[\left((\gamma^k_k)^2-\gamma_{ij}\gamma^{ij}\right)
/(2\epsilon-R_g)\right]^{1/2},\label{lapse}
\end{equation}
where we use the notational shorthand
$\gamma_{ij}=\frac12\dot g_{ij}-X_{(i|j)}$, and then substituting into
(\ref{imco}) gives the Reduced Thin Sandwich Equations (RTSE), for the shift
vector $X^i$ alone. Explicitly,
\begin{equation}\label{rtse}
\left(\sqrt{\frac{2\epsilon-R_g}{(\gamma^n_n)^2-\gamma_{mn}\gamma^{mn}}}
\left(\gamma^{ij}-g^{ij}\gamma^k_k\right)\right)_{|j}=S^i.
\end{equation}

There are two basic conditions we must assume before this equation
can make sense. The first is an apriori restriction on the solution $X^i$
and hence on the data
$(g_{ij},\dot g_{ij},\epsilon,S^i)$, since we regard (\ref{lapse}) as the
definition of $N$ in terms of the data and $X^i$: we must assume
\begin{equation}
N>0\ \ \ {\rm everywhere\ in}\ {\cal M}.\label{bound}
\end{equation}
Physically, this is the superspace requirement that the spacetime slicing
``advance in time''.
Although it is elementary to construct
examples $(g_{ij},\dot g_{ij},N,X^i,\epsilon,S^i)$
satisfying (\ref{enco}), (\ref{imco}), (\ref{excur}) for which $N$
changes sign (by choosing a 1-parameter family of
{\em intersecting} Cauchy surfaces in ${\cal V}$),
such examples will not satisfy (\ref{lapse}), (\ref{rtse}) in the
strict sense since the denominator in (\ref{rtse}) will vanish at some points;
furthermore  they cannot  satisfy the conditions needed for the general
results we describe below, and  they do not satisfy the superspace ``direction
of time'' requirement. The second condition does not have a clear-cut physical
interpretation, but does have a geometric formulation solely in terms of the
given data:
\begin{equation}
2\epsilon-R_g>0\qquad {\rm everywhere\ in}\ {\cal M}.\label{condi}
\end{equation}
In order to minimise technical complications, we shall also assume throughout
that {$\cal M$} is a compact manifold without boundary. It is quite possible
that a more detailed analysis could extend our results, for example, to the
case of asymptotically hyperboloidal data.

The need for the apriori bound (\ref{bound}) means that our main result is
perturbative:
\begin{theo}\label{mtheo1}
If the TSE with data $(g_{ij},\dot g_{ij},\epsilon,S^i)$ admits a solution
$(N,X^i)$ such that the conditions (\ref{bound}) and (\ref{condi}) are
satisfied, and if furthermore, the equation
\begin{equation}
M_{(i|j)}=\mu K_{ij}\label{cond}
\end{equation}
has only the trivial solution $\mu=0$, $M_i=0$, then the TSE with data
$(g_{ij}+\delta g_{ij},\dot g_{ij}+\delta\dot g_{ij},\epsilon+\delta\epsilon,
S^i+\delta S^i)$ sufficiently close to $(g_{ij},\dot g_{ij},\epsilon,S^i)$,
has a unique solution $(N+\delta N,X^i+\delta X^i)$ close to $(N,X^i)$.
\end{theo}

In other words, there is an open neighbourhood of the data
$(g_{ij},\dot g_{ij},\epsilon,S^i)$ in which the TSE is solvable. Our method
of proof is to show that the linearisation of (\ref{rtse}) is
elliptic under suitable conditions, and then apply the Implicit Function
theorem -- it is here that the technical condition (\ref{cond}) arises. Note
that the projections of a spacetime Killing vector satisfy (\ref{cond}).

The reduced equations (\ref{rtse}) can be written formally as
\begin{equation}
F^i(\Psi,X^j)=0,
\end{equation}
where $\Psi$ represents the prescribed fields,
$\Psi=(g_{ij},\dot g_{ij},\epsilon,S^i)$, and $X^j$ is the shift vector.
The linearised thin sandwich operator
$L=\delta F/\delta X$ is given explicitly by
\FL\[
\ \ L^i(M^j)=
\Bigl[{1\over N}\bigl(g^{ij}M^k_{\ |k}-M^{(i|j)}
-{\pi^{ij}\pi_{kl}M^{k|l} \over 2\epsilon - R_g}
\bigr)\Bigr]_{|j},
\]
where $M^j=\delta X^j$ and $\pi^{ij}$ is the (non-density) conjugate momentum,
\begin{equation}
\pi_{ij}=K_{ij}-g_{ij}K^k_k=N^{-1}\left(\gamma_{ij}
-g_{ij}\gamma^k_k\right). \label{momen}
\end{equation}

We now consider properties of the linearised operator $L$; a simple
computation first shows that $L$ is self adjoint. Remarkably, $L$ is
{\em elliptic}
if the conjugate momentum is either positive definite or negative definite:
to see this we calculate the leading order symbol of $L$,
\FL\begin{equation}
\ \ \sigma(L)^i_j(\xi)=\frac1{N}\left(\case1/2\xi^k\xi_k\delta^i_j-
\case1/2\xi^i\xi_j
+{\pi^i_k\pi^l_j\xi^k\xi_l\over 2 \epsilon- R_g}  \right).
\end{equation}
An operator $L$ is elliptic when the symbol
$\sigma(L)$ is invertible for all $(\xi^i)\not=0$. In an
orthonormal frame in which $(\xi^i)=(\xi,0,0)$ we have
\[
\det\left(\sigma(L)(\xi)\right)=\frac{\xi^6}{4N^3(2\epsilon-R_g)}
(\pi_{11})^2,
\]
and hence in general coordinates,
\begin{equation}
\det\left(\sigma(L)(\xi)\right)=\frac1{4N^3(2\epsilon-R_g)}
\xi^k\xi_k(\pi_{ij}\xi^i\xi^j)^2,
\end{equation}
which will be nonzero for all non-zero $\xi^i$ if and only if
$\pi_{ij}$ is a definite matrix.

If condition (\ref{condi}) and the energy constraint (\ref{enco}) hold, then
$\frac12(\pi^i_i)^2-\pi^{ij}\pi_{ij}>0$. Introducing an
orthonormal frame in which $\pi_{ij}$ is diagonal,
$\pi_{ij}=\rm{diag}(\alpha,\beta,\gamma)$ say, this gives
$\frac12(\alpha+\beta+\gamma)^2-(\alpha^2+\beta^2+\gamma^2)>0$, which may be
rearranged as
\[
2\alpha\beta-\case1/2(\alpha+\beta-\gamma)^2>0.
\]
Clearly this requires that $\alpha\beta>0$ and similarly $\beta\gamma>0$,
$\alpha\gamma>0$, which implies that $\alpha$, $\beta$, $\gamma$ all have the
same sign; in other words, $\pi_{ij}$ is definite. We have shown
\begin{propo}
If conditions (\ref{bound}), (\ref{condi}) hold, then the linearised RTS
operator $L$ is self-adjoint and elliptic.
\end{propo}
We note that a similar analysis can be made of the more complicated
reparameterisations of the constraint equations proposed by Komar and
Bergmann \cite{Kom2}, see \cite{Fodor}.

To apply the implicit function theorem, we need to show that $L$ is
surjective, or equivalently, that the null space of $L$ is trivial.
\begin{propo}
Suppose $X^i$ is a solution of (\ref{rtse}) with data $\Psi$ such that
conditions (\ref{bound}), (\ref{condi}) are satisfied. Then $M^i$ is in the
null space of $L$ iff there is a function $\mu$ such that $(M^i,\mu)$
satisfy (\ref{cond}).
\end{propo}
{\em Proof:\ } Integrating $M^iL_i(M^j)$ and dropping divergence terms gives
\[
\int\limits_{\cal M}{1\over N}\biggl(M_{(i|j)}M^{i|j}
-\bigl(M^i_{\ |i}\bigr)^2+
\frac{\left(\pi^{ij}M_{i|j}\right)^2}
     {2\epsilon-R_g}
\biggr) dv=0.
\]
Using the scalar constraint (\ref{enco}), we can rewrite this integral
in the homogeneous form,
\begin{eqnarray}
0=\int\limits_{\cal M}dv\ \frac1{N(2\epsilon-R_g)}\biggl\{
\left(\pi^{ij}M_{i|j}\right)^2+
\ \ \ \ \ \ \ \ \ \ \ \ \ \ \ \ \nonumber\\
\Bigl(\case1/2(\pi^k_k)^2-\pi^{kl}\pi_{kl}
\Bigr)\Bigl(M_{(i|j)}M^{i|j}-\bigl(M^i_{\ |i}\bigr)^2\Bigr)
\biggr\}.\ \ \ \nonumber
\end{eqnarray}
Choosing an orthonormal frame in which the extrinsic curvature is diagonal,
$K_{ij}={\rm diag}(\alpha,\beta,\gamma)$, we have
$0<2\epsilon-R_g=2(\alpha\beta+\alpha\gamma+\beta\gamma)$.
A simple calculation shows that the expression in braces can be written
\FL\begin{eqnarray}
 (\beta+\gamma)^{-2}\Bigl((\beta+\gamma)^2M_{1|1}
+\gamma^2M_{2|2}+\beta^2M_{3|3}\qquad{}\ {}\qquad
\nonumber\\
{}-(\alpha\beta+\alpha\gamma+\beta\gamma)(M_{2|2}+M_{3|3})\Bigr)^2
\qquad{}\quad
\nonumber\\
{}+4(\alpha\beta+\alpha\gamma+\beta\gamma)\Bigl(
(\gamma M_{2|2}-\beta M_{3|3})^2 (\beta+\gamma)^{-2}
\qquad{}\ \qquad{}
\nonumber\\
{}+
(M_{(1|2)})^2+(M_{(1|3)})^2+(M_{(2|3)})^2
\Bigr).\ \qquad\nonumber
\end{eqnarray}
Since $2\epsilon-R_g>0$, this expression is manifestly non-negative and zero
iff $M_{(i|j)}$ is diagonal in this frame and satisfies
$\alpha M_{2|2}=\beta M_{1|1}$, $\alpha M_{3|3}=\gamma M_{1|1}$,
$\beta M_{3|3}=\gamma M_{2|2}$. In this case $M_{(i|j)}$ is proportional to
$K_{ij}$ and $M_{(i|j)}=\mu K_{ij}$ for some function $\mu$. Conversely, if
$(M^i,\mu)$ satisfy (\ref{cond}), then it is easily verified that
$L(M^i)=0$.$\Box$

The equation (\ref{cond}) also arose in the work of  Belasco and
Ohanian\cite{BelOha}, who considered the RTSE as the Euler-Lagrange equations
of the action functional
\FL\[
I = -2\int\limits_{\cal M}\left(\sqrt{\left(\bigl(\gamma^k_k\bigr)^2
-\gamma^{kl}\gamma_{kl}\right)(2\epsilon-R_g)}
+S_iX^i\right)dv.\ \quad {}
\]
They showed
a global form of the previous result\cite{BO-error}:
\begin{propo}
Suppose (\ref{condi}) is satisfied. If $X^i$, $\tilde X^i$ are two solutions of
the RTSE (\ref{rtse}), then there is a positive function $\alpha$ such that
\begin{equation}
\tilde\gamma_{ij}=\alpha\gamma_{ij},\label{gtilde}
\end{equation}
where $\tilde{\gamma}_{ij} = \case1/2\dot{g}_{ij}-\tilde{X}_{(i|j)}$.
Conversely, if $X^i$ is a solution, and $\tilde X^i$ satisfies (\ref{gtilde}),
then $\tilde X^i$ also satisfies (\ref{rtse}).
\end{propo}
It is easily verified that $M^i=\tilde X^i-X^i$ and $\mu=N(1-\alpha)$
satisfy (\ref{cond}) and conversely, if $X^i$ solves (\ref{rtse}) and
$(M^i,\mu)$ solves (\ref{cond}) (and $\mu>N$) then $\tilde X^i=X^i+M^i$
also solves (\ref{rtse}). Since the spacetime vector $(\mu,-M^i)$ satisfies
the spatial part of the Killing equation if (\ref{cond}) holds,
by choosing the lapse/shift $(\mu,-M^i)$ we see that the
underlying data $(g_{ij},K_{ij})$ admits a representation with
 $\dot g_{ij}=0$ (but this choice of lapse will not satisfy
(\ref{bound}) in general).

To formally state our main result, let $H^n(T^i_j)$ denote the set of
tensors of valence $(i,j)$ belonging to the Sobolev space $H^n=W^{n,2}$.
The operator $F(\Psi,X^i)$ defines a smooth mapping
\begin{equation}
F:{\cal H}_n\times H^{n+2}(T^1_0)\rightarrow H^n(T^1_0)
\end{equation}
(${\cal H}_n=H^{n+2}(T^0_2)\times H^{n+1}(T^0_2)\times H^{n+1}(T^0_0)
\times H^n(T^1_0)$), for any $n\geq 2$, assuming that $(\Psi,X^i)$ satisfies
(\ref{bound}), (\ref{condi}). From the Implicit Function Theorem we obtain
\begin{theo}
If $n\geq 2$ and $(\Psi,X^i)\in{\cal H}_n\times H^{n+2}(T^1_0)$ is a solution
of
the RTSE (\ref{rtse}) and satisfies conditions (\ref{bound}), (\ref{condi}),
and if furthermore the equation (\ref{cond}) has only the zero solution, then
there is an open neighbourhood of $\Psi\in{\cal H}_n$ such that the RTSE
(\ref{rtse}) is uniquely solvable for any choice of data $\tilde\Psi$ in this
neighbourhood.
\end{theo}

There is a natural way of generating  $(\Psi,N,X^i)$
satisfying the Thin Sandwich Equations.
Any choice of spacelike hypersurface in an spacetime satisfying the
Einstein equations induces a solution $(g_{ij},K_{ij},\epsilon,S^i)$ of the
constraint equations. Then from
\begin{equation}
\dot g_{ij}=2NK_{ij}+2X_{(i|j)},\label{gdot}
\end{equation}
we see that an arbitrary choice of
shift vector $X^i$ and lapse function $N$ fixes the time derivative
of the spacelike metric $\dot g_{ij}$ along the time flow vector field given
by this shift and lapse.
If $N$ is chosen positive and if $2\epsilon -R_g > 0$, then
 $X^i$ is a solution of (\ref{rtse}).
 In view of
this we can rewrite our main theorem in a more natural form.
\begin{theo}
Suppose  $g_{ij}\in H^{n+2}(T^0_2)$, $K_{ij}\in H^{n+1}(T^0_2)$,
$\epsilon\in H^{n+1}$ and $S^i\in H^n(T^1_0)$ satisfy the constraint
equations (\ref{enco}) and (\ref{imco}), and $2\epsilon-R_g>0$, where $R_g$
is the scalar curvature of $g_{ij}$. For any choice of shift vector
$X^i\in H^{n+2}(T^1_0)$ and positive  lapse function $N\in H^{n+1}$ we define
$\dot g_{ij}$ by (\ref{gdot}). If
the equation $M_{(i|j)}=\mu K_{ij}$ has only the trivial solution, then there
exists a unique continuous map on an open neighbourhood of
$\Psi = (g_{kl},\dot g_{kl},\epsilon,S^k)$ in ${\cal H}_n$, which assigns
$\bar X^i\in H^{n+2}(T^1_0)$ to
$\bar{\Psi}$
such that
$\bar X^k$ is a solution of the reduced thin-sandwich equations with
data $\bar{\Psi}$.
\end{theo}

Finally we  show that there exist reference solutions
$(\Psi,N,X^i)$ satisfying the
conditions of these theorems. If $(\mu,M^i)$ satisfies (\ref{cond}) then
\[
M_{i|j}^{\ \ \ j}+R_{ij}M^j-2(\mu K_{ij})^{|j}+(\mu K^j_j)_{|i}=0,
\]
where $R_{ij}$ is the Ricci tensor of $g_{ij}$. Multiplying by $M^i$,
integrating over ${\cal M}$, and substituting (\ref{cond}) and the energy
constraint (\ref{enco}) gives
\FL\[
\int\limits_{\cal M}
\left\{-M^{i|j}M_{[i|j]}+M^iM^jR_{ij}
-\mu^2(2\epsilon-R_g)\right\}dv = 0.\quad\
\]
If $R_{ij}$ is negative definite, this implies $(\mu,M^i)\equiv(0,0)$ and
thus (\ref{cond}) has only the trivial solution.
Since the $k=-1$ Robertson-Walker spacetimes have spatial slices with
$R_{ij}=-2g_{ij}$, we conclude
\begin{coro}
The Thin Sandwich Equations are uniquely solvable for data $\Psi$ in an
${\cal H}_n$-neighbourhood of the  spatially compactified $k=-1$
 vacuum Robertson-Walker data
$\Psi_0$, for which $({\cal M},g_{ij})$ is a compact hyperbolic 3-manifold,
$\dot g_{ij}=2g_{ij}$, $\epsilon=0$, $S^i=0$, and we use the reference
solution $X^i=0$, $N=1$.
\end{coro}
A similar conclusion obtains for any initial data set $(g_{ij},K_{ij})$
satisfying $R_{ij}<0$ and $\epsilon \ge 0$, provided that the lapse $N$
is chosen strictly positive.

\nonum\section{Discussion}

We have shown that the Thin Sandwich Equations are solvable provided certain
geometric conditions (\ref{bound}), (\ref{condi}) and a technical condition,
related to the absence of Killing vectors, are satisfied. However, since the
linearised equations are not elliptic if $\pi_{ij}$ is indefinite, we do
not expect that the TSE will be well-posed in general, even if $N>0$.
For this reason, the underlying ``superspace'' metaphor must be
considered as mathematically inadequate for the description of the
general Einstein evolution.
However, we note that on the basis of the results presented
here, it is plausible that a suitably restricted form of the
of the full (thick) sandwich conjecture may be valid.

It is useful to relate our thin sandwich results to the
rather satisfactory ADM formulation of the Einstein equations.
The constraint set ${\cal C}={\cal C}(\epsilon,S^i)$, consisting of data
$(g_{ij},K_{ij})$ satisfying (\ref{enco}), (\ref{imco}), forms a natural
(and general) phase space, but unlike $T{\cal S}$, ${\cal C}$ is not easily
parameterised.
We can consider the solution of the TSE as defining a projection
from $(g,\dot{g})$ to $(g,K)\in {\cal C}$, with the fibres of the projection
consisting of classes of data $(g,\dot{g})$ which are equivalent under
change of lapse and shift.
This projection is well-defined on a region in $T{\cal S}$ described by the
two conditions (\ref{bound}), which is an implicit condition on $(g,\dot{g})$,
and (\ref{condi}), which defines the image region in ${\cal C}$ of the
projection,
provided also that the ``Killing'' condition (\ref{cond}) holds.
Thus, solving the TSE provides a parameterisation of that subset
of the full constraint manifold ${\cal C}$ which is determined by the
above restrictions.

The analysis by Arms, Fischer, Marsden and Moncrief \cite{AMM} shows
that the vacuum constraint manifold ${\cal C}$ (on a compact manifold ${\cal
M}$),
has a conical singularity
 when the resulting (vacuum) spacetime admits a Killing vector.
As might have been expected,
in this case (\ref{rtse}) is not solvable in general, since
(\ref{cond}) admits nontrivial solutions.
However the converse is false, since the space of solutions
of (\ref{cond}) is in general larger than the space of
(projected) Killing vectors ---
indeed,  if $K_{ij}\equiv 0$ then $\mu$ in (\ref{cond}) can
be chosen arbitrarily, hence there is an infinite-dimensional
space of solutions of (\ref{cond}) \cite{Fodor}.
The correspondence \cite{Moncrief}
between spacetime Killing vectors and
singularities of the constraint manifold arises from the
elegant relationship between the constraint equations and
the ADM Hamiltonian formulation of the Einstein equations.
The fact that solvability of the TSE requires the more restrictive
condition that (\ref{cond}) have no non-trivial solutions,
indicates that the thin sandwich equations
(and, arguably, the underlying superspace approach)
are not as well-adapted to the Einstein evolution.


\begin{references}
\bibitem[*]{GF-Budapest}Current address:
Central Research Institute for Physics,
Research Institute for Particle and Nuclear Physics,
H--1525 Budapest 114, P.O. Box 49, Hungary
\bibitem{MTW} C. W. Misner, K. S. Thorne and J. A. Wheeler, {\em Gravitation},
(San-Francisco, Freeman, 1973);
R. F. Baierlein, D. H. Sharp and J. A. Wheeler,
{\em Phys. Rev.} {\bf 126}, 1864 (1962)
\bibitem{ChrFra}
D. Christodoulou and M. Francaviglia, in {\em Isolated
Gravitating Systems in General Relativity}, J. Ehlers, Ed. (North-Holland,
New York, 1979), p. 480;
C. M. Pereira, {\em J. Math. Phys.} {\bf 22}, 1064 (1981)
\bibitem{comment} Many writers instead consider superspace to be the
diffeomorphism equivalence classes of metrics. Since the actual
computations are best expressed using the space of metrics, the definition of
${\cal S}$ given here appears to be more natural.
\bibitem{Fodor} G.~Fodor, PhD.~thesis (1993)
\bibitem{ChoBru} Y. Choquet-Bruhat, in {\em Gravitation: An Introduction to
Current Research}, L. Witten, Ed. (New York: Wiley, 1962)
\bibitem{ChBrGe} Y. Choquet-Bruhat and R. P. Geroch,
{\em Commun. Math. Phys.} {\bf 14}, 329 (1969);
T. Hughes, T. Kato and J.  Marsden, {\em Arch. Rat. Mech. Anal.}
{\bf 63}, 273 (1976);
Y. Choquet-Bruhat, D. Christodoulou and M. Francaviglia,
{\em Ann. Inst. H. Poincar\'e}  {\bf 29}, 241 (1978)

\bibitem{Kom2} A. Komar, {\em Phys. Rev. D}, {\bf 4}, 927 (1971);
P. G. Bergmann, {\em Gen. Rel. Grav.} {\bf 2}, 363 (1971)
\bibitem{BelOha} E. P. Belasco and H. C. Ohanian, {\em J. Math. Phys.},
{\bf 10}, 1503 (1969)
\bibitem{BO-error}
{In \cite{BelOha} it is claimed, erroneously,  that
the function $\mu\equiv0$. Counterexamples to this
assertion are given in \cite{Fodor}.}
\bibitem{AMM}
A. E. Fischer and J. E. Marsden, in {\em Isolated
Gravitating Systems in General Relativity}, J. Ehlers, Ed. (North-Holland,
New York, 1979), p322;
J. M. Arms, J. E. Marsden and V. Moncrief, {\em Ann. Phys.}
{\bf 144}, 81 (1982)
\bibitem{Moncrief}
V. Moncrief, {\em J. Math. Phys.} {\bf 16}, 493 (1975)


\end{references}
\end{document}